\documentclass[aps,prx,twocolumn,superscriptaddress,nofootinbib]{revtex4-2}

\usepackage{dsfont}
\usepackage{xcolor,graphicx}
\usepackage{hyperref}
\usepackage{tabulary,graphicx}
\graphicspath{{Images/}}
\usepackage{amsmath,wasysym,amsfonts,amssymb}
\usepackage{mathtools}
\usepackage{hyperref}

\usepackage{mathrsfs}

\usepackage{tikz}
\usetikzlibrary{quantikz}

\DeclareMathOperator*{\tr}{Tr}

\usepackage{xcolor}

\begin{document}

\title{ Variational Adiabatic Gauge Transformation on real quantum hardware for effective low-energy Hamiltonians and accurate diagonalization} 

\author{Laura Gentini} 
\address{Dipartimento di Fisica e Astronomia, Universit\`a di Firenze, I-50019, Sesto Fiorentino (FI), Italy}
\address{INFN, Sezione di Firenze, I-50019, Sesto Fiorentino (FI), Italy}

\author{Alessandro Cuccoli}
\address{Dipartimento di Fisica e Astronomia, Universit\`a di Firenze, I-50019, Sesto Fiorentino (FI), Italy}
\address{INFN, Sezione di Firenze, I-50019, Sesto Fiorentino (FI), Italy}

\author{Leonardo Banchi}
\address{Dipartimento di Fisica e Astronomia, Universit\`a di Firenze, I-50019, Sesto Fiorentino (FI), Italy}
\address{INFN, Sezione di Firenze, I-50019, Sesto Fiorentino (FI), Italy}

\begin{abstract}
	Effective low-energy theories represent powerful theoretical tools to reduce
	the complexity in modeling interacting quantum many-particle systems. However, 
	common theoretical methods rely on perturbation theory, which limits their applicability 
	to weak interactions.  Here we introduce 
	the Variational Adiabatic Gauge Transformation (VAGT), a non-perturbative 
	hybrid quantum algorithm that can use nowadays quantum computers to learn the 
	variational parameters of the unitary circuit that brings the Hamiltonian to either 
	its block-diagonal or full-diagonal form. If a Hamiltonian can be diagonalized via 
	a shallow quantum circuit, then VAGT can learn the optimal parameters using a polynomial number 
	of runs. The accuracy of VAGT is tested trough numerical simulations, as well as simulations 
	on Rigetti and IonQ quantum computers. 
\end{abstract}

\maketitle

\section*{Introduction}

Low-energy approximations permeate many-body physics. Systems as diverse as
cold atoms in optical lattices \cite{duan2003controlling}, solid state spin systems 
\cite{wagner1986unitary}, and even current superconducting quantum computers 
\cite{krantz2019quantum} are accurately described by low energy theories. 
Powerful theoretical methods have been developed to obtain low-energy 
Hamiltonians from perturbative expansions, 
such as the Schrieffer-Wolff transformation \cite{bravyi2011schrieffer},
or non-perturbative methods \cite{burgarth2021eternal}.
However, due to the exponentially large Hilbert space, numerical calculations 
rapidly become unfeasible when the dimensionality of the Hilbert space increases,
while analytical results are limited to toy models. 

Quantum computers and simulators \cite{preskill2018quantum,georgescu2014quantum} are starting to become 
experimentally available, even in the cloud \cite{amazon}. It has been shown that 
quantum computers can accurately approximate the ground state of many-particle 
systems \cite{peruzzo2014variational,kandala2017hardware,motta2020determining}, estimate 
molecular energies \cite{o2016scalable}, molecular docking configurations
\cite{banchi2020molecular}, and even some excited states 
\cite{higgott2019variational}. 
One of the main challenges in quantum simulation is computing the dynamics of
quantum many-particle systems without having to resort to exact diagonalization or
conventional perturbation theory. Algorithms based on the Suzuki-Trotter 
decomposition \cite{georgescu2014quantum,csahinouglu2021hamiltonian} have been adapted to better
exploit the capabilities of current noisy hardware 
\cite{li2017efficient}, while accurate evolutions for longer times can be obtained 
using variational fast-forwarding \cite{cirstoiu2020variational} 
or Hamiltonian diagonalization \cite{commeau2020variational, JonesE19} methods. 

Here we introduce a hybrid variational quantum algorithm for either block- 
or full-diagonalization of $N$ qubit Hamiltonians, where complex calculations in exponentially large Hilbert
spaces are performed on a quantum hardware while, when some assumptions are met,
the classical part of the algorithm scales polynomially in $N$. 
Our algorithm provides an efficient way of finding a variational circuit that brings the 
Hamiltonian to the desired diagonal or block-diagonal form, which can be used to extract low-energy 
interactions or estimate quantum dynamics as in fast forwarding. 
Our method is based on recent advances \cite{PhysRevB.101.014302, SelsE3909, HatomuraT21, SugiuraC21, KOLODRUBETZ20171} in the context of adiabatic gauge potentials (AGPs), which are
infinitesimal generators of a unitary transformation diagonalizing a given
Hamiltonian. The AGP is non-perturbative,  it can recover Schrieffer-Wolff transformation in the 
perturbative limit, and it is tightly connected to the Wegner Hamiltonian flow 
\cite{wegner1994flow}, also called similarity renormalization group 
\cite{anderson2008block}. We define the Variational Adiabatic Gauge Transformation (VAGT), 
as a variational quantum circuit approximation to the unitary generated by the AGP, and 
show that the parameters of that transformation can be efficiently trained on 
noisy-intermediate-scale quantum (NISQ) devices \cite{preskill2018quantum}.  

We show that VAGT yields an accurate block diagonalization with few variational parameters (low depth)
when the Hamiltonian has some energy separated or symmetry separated blocks, and 
full-diagonalization when the number of parameters increases. 
Finally, the feasibility of our method on current noisy quantum hardware is 
tested with experiments on the Rigetti Aspen-9 and IonQ 11-qubit 
quantum processors.

\section*{Results}

We focus on the diagonalization, or block-diagonalization, of 
a Hamiltonian $H$, assuming that there is another Hamiltonian 
$H_0$ whose eigenvalues and eigenstates are known, and possibly 
easy to prepare on a quantum device. We thus split $H$ as
\begin{equation}
\label{eq:H_lambdadef}
H\equiv H_\lambda = H_0+\lambda V, 
\end{equation}
where $\lambda V=H-H_0$, and $\lambda$ models the strength of the correction.
 A good approximation of the ground state of $H$ can be
prepared thanks to the adiabatic theorem \cite{albash2018adiabatic}, 
by starting from the
ground state $\ket{g_0}$ of $H_0$ and then slowly increasing the
interaction strength $\lambda$. For an evolution time
$T$, the approximate ground state is obtained as $ \ket{g}=\mathcal T
\exp\left({-}i \int_0^T H_{\mu(t)} dt\right)\ket{g_0} $, where
$\mu(t)$ is a function, typically linear in $t$,
satisfying $\mu(0)=0$ and $\mu(T)=\lambda$.
Such adiabatic preparation of the ground state is accurate and efficient when the ground state of $H_\mu$ is non-degenerate and well separated from the excited 
states for all  $\mu\in[0,\lambda]$. 

A generalization of the adiabatic ground state preparation is given 
by the adiabatic gauge potential 
\cite{PhysRevB.101.014302,SelsE3909, ClaeysP19}, which 
defines the infinitesimal generators of a unitary transformation
that allows the estimation of more eigenvalues, in some cases even
performing full-diagonalization -- see also 
Appendix~\ref{sec:AGP} for more details. 
Consider some infinitesimal generators $A_\mu$ 
for $\mu\in[0,\lambda]$, set the unitary 
\begin{equation}
U_\mu = \mathcal T_\nu\exp\left({-}i\int_0^\mu A_\nu d\nu\right) ~ \Longleftrightarrow 
~ A_{\mu} = i(\partial_\mu U_\mu)U_\mu^\dagger,
\label{UA}
\end{equation}
and the rotated Hamiltonian 
\begin{align}
	\tilde H_\mu &:= U_\mu^\dagger H_\mu U_\mu, 
	\label{Htilde}
\end{align}
with $\mathcal T_\nu$ denoting the ordering with respect to $\nu$. 
\begin{figure}[t] \centering
	\includegraphics[width=0.8\linewidth]{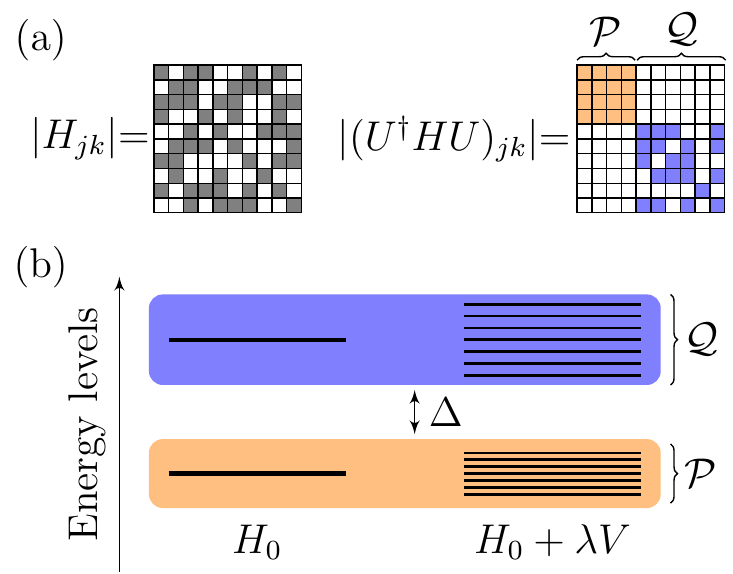}
	\caption{Pictorial representation of the action of the adiabatic
		gauge transformation $U$, which performs a block-diagonalization
		of $H=H_0+\lambda V$ separating the low-energy ($\mathcal{P}$)
		and high-energy ($\mathcal{Q}$) Hilbert spaces. 
		(a) Example matrix elements of $H$ before (left) and
		after (right) the action of $U$.  
		(b) Example structure of $H_0$ and $V$: the term $V$ breaks 
		the symmetries of $H_0$ and splits its degenerate eigenvalues,
		still maintaining them in two separate subspaces with suitably 
		large energy separation $\Delta$. 
	} \label{fig:cartoon}
\end{figure}
Depending on the problem, we want to reach a diagonal 
or block-diagonal $\tilde H_\mu$ at the end of the evolution, when
$\mu=\lambda$ (see Fig.~\ref{fig:cartoon}(a)). 
We assume for the sake of simplicity that $H_0$ has two blocks which, in the easiest case, correspond to two degenerate eigenvalues $h_{\mathcal{P}}$ and $h_{\mathcal{Q}}$ of $H_0$, 
with $h_{\mathcal{P}}\ll h_{\mathcal{Q}}$, 
that are possibly split by the term $\lambda V$, as in
Fig.~\ref{fig:cartoon}(b) 
--
the extension to Hamiltonians with more blocks is straightforward.
In such case, we call $P$ and $Q$, 
respectively, the projectors on the low-energy and high-energy 
sectors, but in general $P$ and $Q$ can also be projectors on symmetry sectors of the Hamiltonian $H_0$. 
For block-diagonalization we should impose 
that all the elements of the off-diagonal blocks are zero, i.e. 
$P\tilde H_\mu Q=0$.  Differentiating such equation with respect to $\mu$, we get
\begin{align}
	P U_\mu^\dagger G_\mu U_\mu Q & =0,
																			 &
	G_\mu &:= V+i[A_\mu,H_\mu].
	\label{PGQ}
\end{align}
The latter equation can also be written as 
$[U_\mu^\dagger G_\mu U_\mu,H_0]=0$
when $H_0=h_{\mathcal{P}} P + h_{\mathcal{Q}} Q$ 
has only two degenerate eigenvalues, as in Fig.~\ref{fig:cartoon}. 

The adiabatic gauge potential  is a particular choice of $A_\mu$
that satisfies the operator equation 
\begin{equation}
[G_\mu ,H_\mu]=0,
\label{C}
\end{equation}
see Appendix~\ref{sec:AGP} for more details. Such equation is stronger than 
Eq.~\eqref{PGQ} and, when exactly satisfied, the resulting 
$\tilde H_\mu$ has no off-diagonal elements. 
Approximations of the above exact solution
were proposed in \cite{PhysRevB.101.014302,SelsE3909, SaberiO14, HartmannL19, PassarellaC20, WurtzP20}, 
based on a variational approximation of the $A_\mu$, with optimal 
parameters obtained by variationally minimising, on a classical computer, either 
$\| [G_\mu ,H_\mu] \|$ 
or $\| G_\mu\|$, where $\|\cdot\|$ 
is the Hilbert-Schmidt norm. With some assumptions, such a variational
approximation of the AGP is efficient in suppressing 
matrix elements between states that belong to different energy sectors or that
are well-separate in the basis defined by the symmetries of 
$H_0$, effectively resulting in a 
block diagonalization of the Hamiltonian.
Therefore, the AGP can also be applied when low-energy and high-energy sectors 
are {\it a priori} unknown.

In this work we propose a different variational approach, whose parameters 
can be optimized in NISQ hardware. 
Taking inspiration from the success of hybrid variational 
quantum algorithms \cite{yuan2019theory, GentiniC20}, we consider
a variational quantum circuit ansatz for the 
Adiabatic Gauge Transformation (AGT) $U_\mu$ defined in Eq.~\eqref{UA}, namely 
we write
\begin{equation}
\label{Ualpha}
U_\mu(\alpha)= U_0 \prod_{\ell=1}^L e^{-i \alpha^\ell_{\mu} B^\ell},
\end{equation}
where $\alpha^\ell_{\mu}$ are variational parameters, 
$L$ is the number of layers in the circuit ansatz, and $B^\ell$
are local operators. Given the available gates in current 
quantum hardware we choose $B^\ell$ such that 
$e^{i \alpha B^\ell}$ is either a one- or two-qubit gate. 
If the Hamiltonian $H_0$ is diagonal in the chosen basis, then $U_0=\openone$,
otherwise we assume that $U_0$ may be efficiently expressed as a known
quantum circuit. 
Each parameter $\alpha^\ell_\mu$ is a continuous function of the
running parameter $\mu \in [0,\lambda]$. By dividing such
interval in $T$ steps $\delta\mu$ we create a
discrete set of $T$ values for $\mu$: 
\begin{align}
	\mu \in [0,\lambda]&\rightarrow \{ \mu_t\} _{t=1}^T,  & 
	\mu_t & = t \delta \mu, & t &\in \mathbb{N} .
\end{align} 
As a result we now have a discrete set of $LT$ variational parameters $\{\alpha^\ell_t\}$.
Within precision $\delta\mu$ the potential $A_{\mu}= i(\partial_\mu U_\mu)U_\mu^\dagger $ at step $t$
can be approximated via finite differences as 
\begin{equation}
	A_{\mu_{t}} 
	\simeq \sum_{\ell=1}^{L} \frac{\alpha^\ell_{t+1}-\alpha^\ell_{t}}{\delta \mu}O^\ell_{t}, 
\end{equation}
where 
$	O^\ell_{t} :=U^\ell_t B^\ell (U^\ell_t)^{\dagger}  $,
	$U^\ell_{t} :=U_0 \prod_{k<\ell}^{\rightarrow} e^{-i \alpha^k_{t} B^k} $.
We set $\alpha^\ell_0=0$ at step $t=0$ for
$1\leq\ell\leq L$ and, starting from this initial configuration, 
we iteratively impose either Eq.~\eqref{PGQ} or 
\eqref{C}, to get firstly $\alpha^\ell_1$ and then the optimal
parameters $\alpha^\ell_t$ at all steps $t$. 
More precisely, as we will clarify in the next sections,  setting $
\beta^{\ell}_t = \frac{(\alpha^\ell_{t+1}-\alpha^\ell_{t})}{\delta \mu }$ 
those equations can be written as $X_t\cdot \beta_t=Y_t$, for 
some operators $X_t$ and $Y_t$. Calling 
$\tilde \beta_t$ the solution of such operator equation we 
get the gradient-like update rule 
\begin{equation}
\alpha_{t+1}^\ell = \alpha_{t}^\ell + \tilde \beta^\ell_t
\delta\mu,
\label{gradient}
\end{equation}
where all $\tilde \beta^\ell_t$ are obtained by classical post-processing 
of quantum measurement results. We notice that, although Eq.~\eqref{gradient}
resembles a gradient ascent update rule, it was obtained from a completely 
different route. 
Parametric quantum circuits like the one in Eq.~\eqref{Ualpha} can give rise 
to barren plateau in the cost function landscape \cite{mcclean2018barren,arrasmith2020effect,cerezo2021cost} 
when the parameters are randomly initialized or for global cost functions. 
However, in the VAGT algorithm the cost function is local and 
all the parameters are initialized to zero and then evolved to the optimal values, 
a strategy that has been found to address the barren plateau problem \cite{grant2019initialization,cerezo2021cost}. 
An update rule similar to Eq.~\eqref{gradient}, namely based on the solution of linear system of equations 
with coefficients estimated via quantum hardware, was discussed in the context of the quantum 
imaginary time evolution algorithm \cite{mcardle2019variational,yuan2019theory}, but the resulting 
circuits are entirely different. 
In the following sections we will study 
different applications that can be done efficiently on a quantum hardware. 

\subsection*{Variational Quantum Adiabatic Gauge Transformation Algorithm}

Since $H$ acts on a Hilbert space whose dimension exponentially 
increases with the number of qubits, in general the  diagonalization or 
block diagonalization of the Hamiltonian is exponentially hard. Here we show
that, provided $H$ can be accurately diagonalized by a shallow circuit, such 
diagonalizing unitary can be found in polynomial time using a hybrid 
quantum-classical algorithm that can be run on nowadays NISQ devices. 
Our algorithm is based on the minimization of the norm $\|G_\mu\|$ that,
as we show in Appendix~\ref{sec:analytic}, is equivalent to the solution of 
the linear system $\sum_{\ell} X^{l,\ell}_t \beta^\ell_t = b^{l}_t$ of $L$ equations,
from which we can update the variational parameters following \eqref{gradient}. 
In order to define a quantum circuit to measure the coefficients 
$X^{l,\ell}_t$ and $b^l_t$ in a quantum computer, we first expand $V$ and $H_{\mu_t}$ 
in term of Pauli operators
\begin{equation}
	V=\sum_{j} v_j \sigma_j \quad \quad H_{\mu_t}=\sum_{j} h_{jt} \sigma_j \, ,
	\label{decompose}
\end{equation}
where $\sigma_j \equiv  \sigma^{k^j_1}\otimes \cdots \otimes \sigma^{k^j_N}$
are strings of Pauli operators, $k_i^j \in\{0,x,y,z\}$ and, in principle, the
sum index runs up to $4^N$ where $N$ is the number of qubits. 
However, in most physical relevant cases the Hamiltonian only contains a limited
number of terms, so most coefficients $h_{jt}$ and $v_j$ are null. The
quantum algorithm we are about to define does not require to run a quantum
circuit corresponding to such zero coefficients. 
In Eq.~\eqref{decompose} we also have dropped the dependence on $\mu_t$ of $h_j$ to simplify notation. 

\begin{figure}[t!]
	\centering
 	\includegraphics[width=0.99\linewidth]{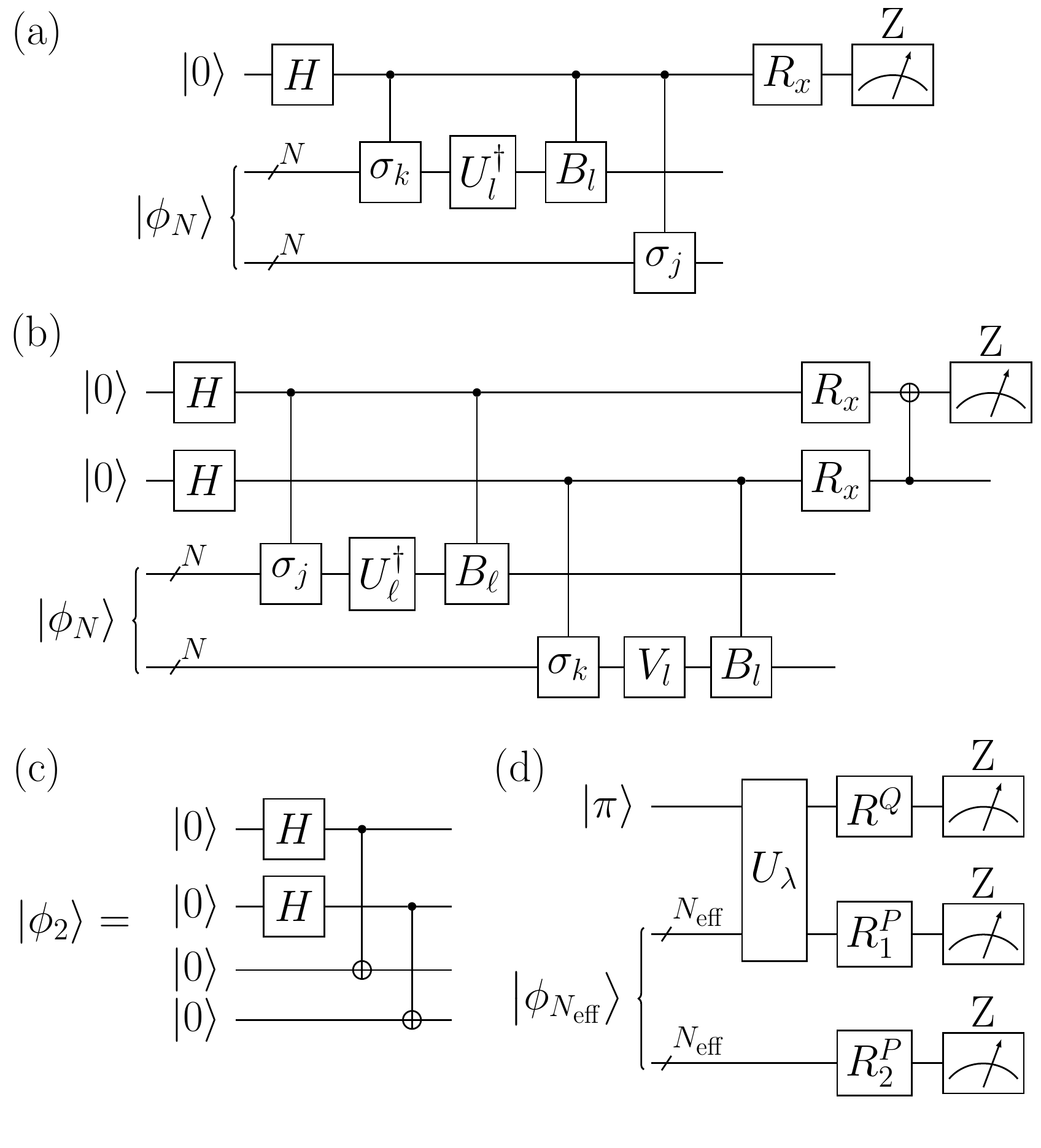}
	\caption{Suitable quantum circuits to evaluate the VAGT trace operations, appearing
		in Eqs.~\eqref{termb}-\eqref{termx}: (a) and (b) show, respectively, the quantum circuits 
		to measure the quantities appearing in $b_t^\ell$ and $X_t^{l, \ell}$;
		(c) sample
		circuit to generate the maximally entangled input states $\ket{\phi_N}$ when $N=2$ -- 
		see also the definition in Eq.~\eqref{eq:phiN}; 
		(d) circuit to measure the expectation values in Eq.~\eqref{heffexpval}, 
		where a product of single qubit rotations $R_j$ is used to transform 
		$\sigma_k\otimes\tilde\sigma_j$ into a product of $Z$ measurements. 
}
	\label{fig:circuits}
\end{figure}

We call {\it computational basis} the basis in which all the Pauli operators $\sigma_k^z$ are diagonal.
In terms of such coefficients we find
\begin{align}
	b_t^l&= -\sum_{j, k} v_j h_{ kt} \text{Tr}(\sigma_j\, i[U_t^l B^l U^{l\dagger}_t, \sigma_k]),
	\label{termb}
	\\
	X_t^{l, \ell} &= \sum_{j, k} h_{ jt} h_{ kt} \text{Tr}(i[U_t^\ell B^\ell
	 U_{t}^{\ell\dagger}, \sigma_j] \,  i[U_t^{l} B^{l}
	 U_{t}^{l\dagger}, \sigma_k]),
	\label{termx}
\end{align}
where the sums are restricted to non-null values of $h_{kt}$ and $v_j$. 
The detailed derivation of the last equations is reported in Appendix~\ref{sec:circuits}, where we also show how such quantities can be estimated on a quantum computer using the circuits given in Fig.~\ref{fig:circuits}, 
where $\ket{\phi_N}$ is the maximally entangled state that can be constructed using $\mathcal O(N)$ operations 
as in Fig.~\ref{fig:circuits}(c). Therefore, the number of operations in each circuit is at most $\mathcal O(N+L)$. 

The scaling efficiency of the method depends on the connectivity of the
Hamiltonians $H_0$ and $V$, namely on the number
of terms $N_V$ and $N_H$ associated to a non-null coefficient in the two quantities in Eq.~\eqref{decompose}. Suppose that 
$\max(N_V,N_H)=\mathcal O(N^\gamma)$: for instance, if the Hamiltonians contain just single-qubit terms, then $\gamma=1$; for nearest neighbour interactions 
$\gamma=1$, too; on the other hand, $\gamma=2$ if $H$ or  $V$ contain all possible 
two qubit interactions.
For each step $t$ the number of circuits needed to evaluate all terms \eqref{termb} and \eqref{termx} 
is respectively 
	$N_b =  \mathcal{O}(N^{2\gamma} L)$,
	$N_X = \mathcal{O}(N^{2\gamma} L^2)$, 
where $L$ is the number of layers in Eq.~\eqref{Ualpha}. 
Therefore, the number of measurements 
to be performed on the quantum device to calculate 
all variational parameters through Eq.~\eqref{gradient} is 
\begin{equation}
	\mathcal O (N^{2\gamma} L^2T),
\end{equation}
while the solution of all linear systems of equations is at most 
$\mathcal O(L^3)$ for each step $t$. 
Therefore, for shallow circuits with $L={\rm poly}(N)$, both 
the algorithmic and measurement complexities scale polynomially in the number of qubits $N$ and linearly 
in the number of steps $T$.

\subsection*{Low-energy approximation }

 When well-defined energy sectors exist for the problem at hand, 
as in Fig.~\ref{fig:cartoon}, the projected block diagonalized Hamiltonian
$ P\tilde{H}P$ can be interpreted as a low energy effective Hamiltonian, that
can be useful in the context of many-body physics, where the original
Hamiltonian may be unmanageable for many purposes, such as calculating dynamics.
Assume that $P$ is known, and that the low-energy block can be expanded in the Pauli basis as 
 \begin{equation}
 H^{\rm eff} = P\tilde H P = \sum_{\tilde j} h^{\rm eff}_{\tilde j} \tilde\sigma_{\tilde j}, 
 \label{eq.heff}
 \end{equation}
where $\tilde \sigma_{\tilde j}$ are Pauli operators acting on $N^{\rm eff}$ qubits,
then the expansion coefficients $h^{\rm eff}_{\tilde j}$ can be obtained using a simple 
quantum circuit. Indeed, using the decomposition \eqref{decompose} we get
\begin{equation}
	h^{\rm eff}_{\tilde j} 
	=  \sum_{j}\frac{h_{jT}}{2^{N^{\rm eff}}} \tr[U_\lambda^\dagger \sigma_{j} U_\lambda 
	P\tilde \sigma_{\tilde j} P],
\end{equation}
and such coefficients can be evaluated in-hardware using a simple circuit like the one in Fig.~\ref{fig:circuits}.
Indeed, suppose that $P=\openone_{N^{\rm eff}}\otimes\ket{\pi}_{N-N^{\rm eff}}\!\bra{\pi}$ and that $\tilde \sigma_{\tilde j}$ 
nontrivially acts only in the ${\mathcal{P}}$ subspace, then we may write 
\begin{equation}
	\frac{	\tr[U_\lambda^\dagger \sigma_{j} U_\lambda 	P\tilde \sigma_{\tilde j} P]}
	{ 2^{N_{\rm eff}}}= \bra{\pi\phi_{N_{\rm eff}}}U_\lambda^\dagger
	\sigma_j U_\lambda \otimes \tilde \sigma_{\tilde j}^* 
\ket{\pi\phi_{N_{\rm eff}}},
\label{heffexpval}
\end{equation}
which can be measured using the circuit shown in Fig.~\ref{fig:circuits}(d). 
In the above equations $U_\lambda\equiv U_\lambda(\alpha_{\rm opt})$ and
$\alpha_{\rm opt}$ are the optimal parameters 
obtained at the end of the iteration, namely with $t=T$.  Therefore, 
provided that the number $N_P$ of non-null expansion coefficients $h^{\rm
eff}_{\tilde j}$ in Eq.~(\ref{eq.heff}) is suitably small, the effective
Hamiltonian can be efficiently obtained. 

\begin{figure}
    \centering
    \includegraphics[width=.45\textwidth]{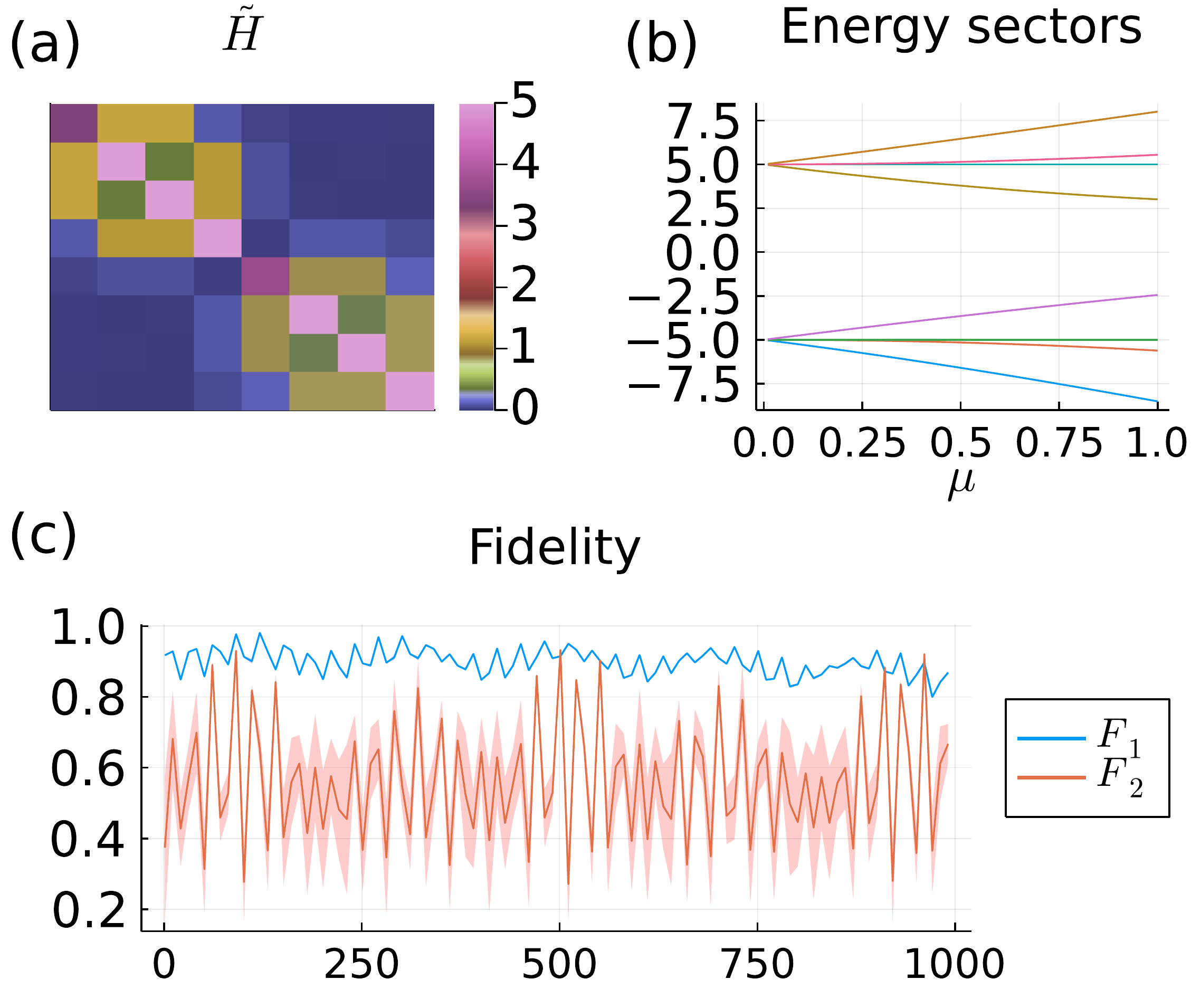}
		\caption{Results obtained from the model Hamiltonian~\eqref{eq:lowH} via
			numerical simulation with $\lambda=1$, $h=-5$, $T=100$ and $L=36$.
			(a) Absolute value of transformed Hamiltonian components $\vert
			\tilde{H}\vert_{jk}$ in the eigenbasis of $H_0$.
			(b) Energy sectors
			defined by the Hamiltonian \ref{eq:lowH} as $\mu$ is running from $0$ to
			$\lambda=1$.
			(c) 
			State fidelity $F_1$ between the time evolved state 
			and the one obtained from the effective model, 
			and fidelity $F_2$ between the evolved and initial states, as defined in 
			Eq.~\eqref{eq:fidelity}, for different random initial states. 
			Solid lines represent the mean value and coloured regions the 95$\%$ confidence interval. 
    }
    \label{fig:low2}
\end{figure}

We test our framework using the following three-qubit model Hamiltonian
\begin{equation}
    H=h\sigma^z_3 +\lambda \bigl[ ( \vec{\sigma}_1 \cdot \vec{\sigma}_3 +  \vec{\sigma}_2 \cdot \vec{\sigma}_3 ) -  (\sigma^x_1 + \sigma^x_2) \bigr] 
\label{eq:lowH}
\end{equation}
with $h = -5$. 
For $\lambda =0$ the Hamiltonian is diagonal in the computational basis with
only two degenerate, well separated energy levels. As $\lambda$ grows, the 
off-diagonal part of $H$ is designed to completely remove the degeneracy, while keeping the levels in two
separate subspaces, as showed in fig \ref{fig:low2}(b). In the Hamiltonian Eq.~\ref{eq:lowH}, qubits $1$ and $2$ do not
interact directly, but can effectively communicate via qubit $3$.
Qubit 3 is the one that define energy sectors, so we can easily
identify the low-energy projector $P$ in the computational basis as 
\begin{equation}
	P = \openone_{12}\otimes \ket{0}_3\!\bra{0} ,
\end{equation}
where $\ket{0}$ is the eigenstate of $\sigma_3^z$ with eigenvalue +1. This
means that, at the end of the process, one can obtain a effective low energy
Hamiltonian that couples qubits 1 and 2, with interactions mediated by qubit 3
without having to take in account its evolution at all. Even if this is just a
toy model, it is reminiscent of quantum communication schemes
\cite{bose2003quantum,banchi2010optimal,banchi2017gating}: if 
qubit 3 is replaced by a multi-qubit communication channel, 
forming for example a qubit chain, then this method can
be used to find an effective Hamiltonian for the sender and receiver
qubits only \cite{wojcik2007multiuser}. 

In Fig.~\ref{fig:low2} we  present  the  results  obtained  with $\lambda=1$ and 
$T=100$. We use a variational ansatz composed by two blocks of layers: the first
one is made by three layers of single qubit rotations around the $x$, $y$ and $z$ axis respectively for each qubit; the second one is made of parametrized two qubit gates,
$\sigma^x \otimes \sigma^x$, $\sigma^y \otimes
\sigma^y$ and $\sigma^z \otimes \sigma^z$, for different pairs of qubits. Since the
Hamiltonian Eq.~\eqref{eq:lowH} is symmetric with respect to the exchange of
qubits $1$ and $2$, we employed a symmetric ansatz where each operator $B^\ell$ in 
Eq.~\eqref{Ualpha} satisfies $[B^\ell,S_{12}]=0$, being $S_{12}$ the swap operator.
Considering such symmetry and alternating and repeating each block three times,
we get $L=36$ free parameters. 
Fig.~\ref{fig:low2}(a) shows the absolute value of transformed
Hamiltonian components $\vert \tilde{H}_{jk} \vert$, where the block structure due
to energy sectors (Fig.~\ref{fig:low2}(b)) is clearly visible. 
The resulting effective interaction between qubits 1 and 2 
is 
\begin{align}\label{eq:Heff}
    H^{\rm eff} &\simeq -1.1\, (\sigma_1^x + \sigma_2^x) + 1.0\,(\sigma^z_1 +\sigma^z_1)+ 
	\\\nonumber
	&-0.2\, (\sigma_1^x\sigma^x_2 
	+ \sigma_1^y\sigma^y_2)  
	 + 0.1\, (\sigma_1^y\sigma^z_2 + \sigma_1^z\sigma^y_2 ),
\end{align}
where only the terms larger than 0.1 have been shown, the full Hamiltonian can be found 
in Appendix~\ref{s:Heff}.
Fig.~\ref{fig:low2}(c) shows the state fidelity $F_1$ between the time evolved state according 
to the full Hamiltonian \eqref{eq:lowH} and the one obtained from the effective model, together with
the fidelity $F_2$ between the evolved and initial states 
\begin{align}
	\begin{aligned}
	F_1(t) &= 	
	\bra{\psi_{12}^{\rm eff}(t)}\rho_{12}(t) \ket{\psi_{12}^{\rm eff}(t)},
				 \\
	F_2(t) &= 	
	\bra{\psi_{12}^{\rm eff}(0)}\rho_{12}(t) \ket{\psi_{12}^{\rm eff}(0)},
	\end{aligned}
	\label{eq:fidelity}
\end{align}
where $\rho_{12}(t)=\tr_3 \ket{\psi_{123}(t)}\!  \bra{\psi_{123}(t)}$, 
\begin{align}
	\ket{\psi_{123}(t)} &= e^{-it  H_{\lambda}}\ket{\xi_{12},0_3}, 
											&
\ket{\psi^{\rm eff}_{12}(t)} &= e^{-it H^{\rm eff}}\ket{\xi_{12}},
\nonumber
\end{align}
$\ket{\xi_{12}}$ are randomly generated two qubit state, 
and $t$ is varied from $1$ to $1000$. 
As shown Fig~\ref{fig:low2}(c), $F_2$ displays a non-trivial behaviour, signaling a non-trivial dynamics. 
Since $F_1\sim1$ for $t\leq 1000$, 
such dynamics is accurately reproduced by the effective model for remarkably long times.


\subsection*{Block diagonalization}
We now study 
the performance of the VAGT algorithm with symmetry
defined blocks, by focusing on the following spin chain Hamiltonian with open boundary conditions 
\begin{equation}
\label{H_sc}
H_\lambda=\sum_{i=1}^{N-1} \bigl( \sigma^x_i\sigma^x_{i+1}
+\sigma^y_i\sigma^y_{i+1} + h \sigma^z_i \bigr) + \lambda \sum_{i=1}^N
\sigma^x_i, 
\end{equation}
where $N$ is the number of qubits, while $h$ and $\lambda$ are, respectively, the
transverse and longitudinal fields. 
For $\lambda=0$ the model is exactly solvable, so eigenvalues and eigenvectors
can be found in $\mathcal O({\rm poly}(N))$ time. For this Hamiltonian the blocks 
can be identified as different magnetization sectors, as $H_0$ commutes with $\sum_i \sigma^z_i$, 
without necessarily being far away in energy. 

\begin{figure}
    \centering
    \includegraphics[width=.45\textwidth]{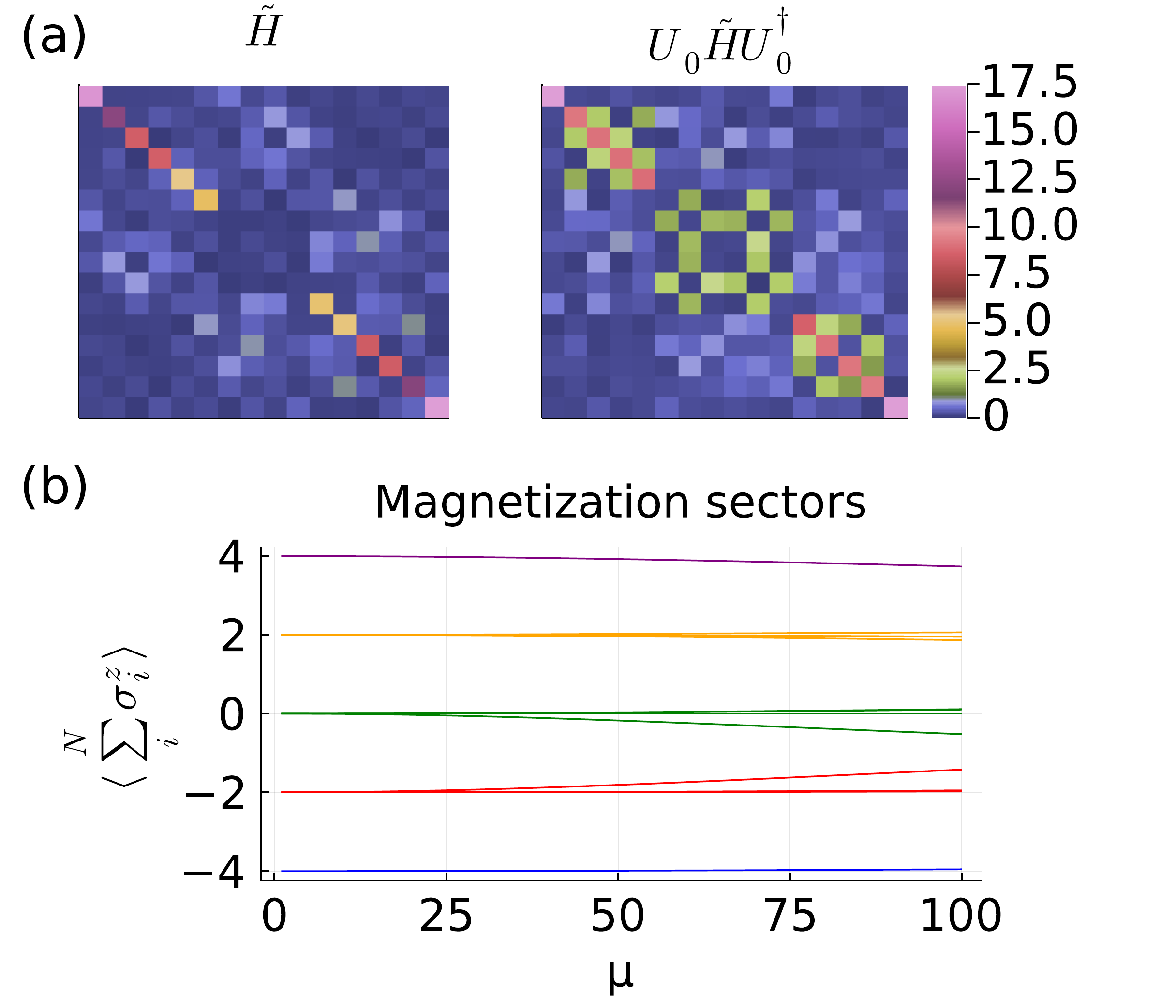}
		\caption{(a) Results obtained from Hamiltonian Eq.\eqref{H_sc}, for
			$N=4, \lambda=1, h=4.5, T=100$ and $L=140$. The transformed Hamiltonian
			$\tilde{H}$ is showed both in the eigenbasis of $H_0$  (left) and in the 
			magnetization eigenbasis of $\sigma_i^z$ (right).
			(b) Magnetization sectors naturally defined by the symmetries of $H_0$ are split for non-zero $\mu$. 
			Values are obtained by analytically
			computing the expectation value of $\sum_i^N \sigma^z_i$ on the eigenvectors of
			$H_\mu$ from Eq. \eqref{H_sc}, for $\mu \in [0,\lambda]$ and $\lambda =1$.
		}
		\label{fig:res4main}
\end{figure}
In Fig.~\ref{fig:res4main} we present the results obtained 
 with $N=4$, $h=4.5$, $\lambda=1$ and $T=100$ by numerical simulation of the
 quantum circuits. The VAGT in Eq.~\refeq{Ualpha} is composed of two blocks of layers, 
 the first block contains two layers of parametrized single qubit rotation gates around the 
 $x$ and $y$ axes, while the second block contains two layers of 
 parametrized two qubit gates. For the latter we choose only nearest-neighbour
 $\sigma^y \otimes \sigma^y$ and $\sigma^z \otimes \sigma^z$ interactions, and we
 alternate and repeat both blocks of layers ten times, resulting in $L=140$. Even if such ansatz is obviously not universal for a 4-qubit system,  
the small off-diagonal terms at the end of the optimization, as shown in the left panel of Fig.~\ref{fig:res4main}(a),
confirm the validity of our algorithm. 
The solution can also be improved by using deeper circuits and finer slicing, i.e.
higher $T$. 
In Fig.~\ref{fig:res4main}(a), we also show (right panel) the transformed
Hamiltonian in the magnetization basis, where the block structure associated with the
different magnetization
sectors defined by the original symmetry is clearly apparent.

\subsection*{Implementation on NISQ devices}
We now discuss the implementation of our algorithm on real quantum hardwares,
the Rigetti Aspen-9 quantum processor with 31 qubits, and IonQ quantum processor with 11 qubits, 
that we access through the cloud-based Amazon Braket service~\cite{amazon}. In order to simplify 
the experiment, we focus on two-qubit Hamiltonians, as in such case, as shown 
in Appendix~\ref{sec:savingmoney}, we can fully exploit some specific properties to minimize the number of gates employed, and accordingly the simulation cost. 
With this simplification, valid for $N=2$, each circuit requires at most 5 qubits. 
In order to fully exploit Aspen-9's 31 qubits and reduce cost, we run 4
different experiments in parallel, still guaranteeing the presence of one or
two ``garbage'' qubits between different experiments to reduce possible cross talk. 
On the IonQ's 11-qubits hardware we run instead 2 experiments in parallel. In numerical 
simulations the full circuits shown in Fig.~\ref{fig:circuits} is implemented.
Using a universal variational ansatz the circuit depth is $L=15$, but lower depths
are possible by using an ansatz suitably designed for the specific Hamiltonian problem at hand. 
For this purpose, we choose to test our method on quantum
hardware with the highly non-diagonal Hamiltonian defined below: 
\begin{align}
\label{eq:H0_nondiag_mt}
H_0 =&  \sigma^z_1 + \sigma^z_2,  \\\nonumber
V =& v_1 \sigma^x_1 + v_2 \sigma^x_2 + v_3 \sigma^y_1 + v_4 \sigma^y_2 + 
v_5 \sigma^x_1 \sigma^x_2 + \\ &+ v_6 \sigma^x_1 \sigma^y_2 
+ v_7 \sigma^y_1 \sigma^x_2 + v_8
\sigma^y_1  \sigma^y_2 , 
\label{eq:V_ops_mt}
\end{align}
with $\lambda=1 $ and where the $v_k$ coefficients are randomly chosen between $0$ and $1$.
Due to the lack of well defined sectors, whether they are defined by energy, magnetization or other physical quantities, we do not expect block-diagonalization in any basis, though thanks to the universal variational ansatz, we can expect full diagonalization in the 
computational basis.
\begin{figure}[t]
	\centering
\includegraphics[width=0.99\linewidth]{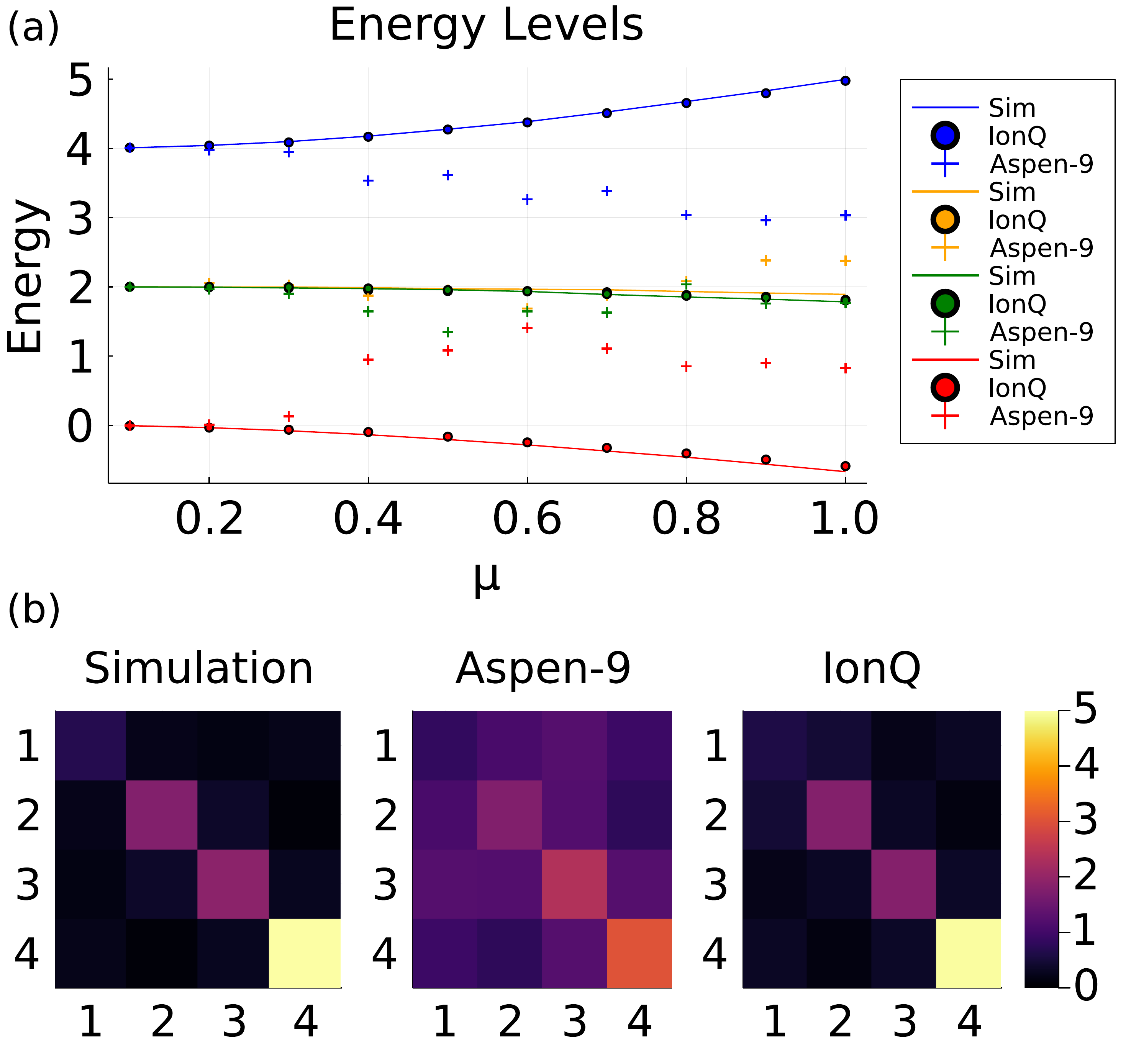}
	\caption{(a) Energy levels for Hamiltonian Eqs.~\eqref{eq:H0_nondiag_mt} and \eqref{eq:V_ops_mt}, 
		as a function of $\mu \in [0... \lambda]$ with $\lambda=1$. 
	(b) Absolute value of the transformed Hamiltonan components $|\tilde H|_{jk}$, as obtained with 
	numerical simulations, and simulations on Rigetti's or IonQ's hardwares. All simulations were performed 
	with $T=10$ discretization steps and $S=100$ shots per measurement. 
	}%
	\label{fig:hardware}
\end{figure}

%
%
%

In Fig.~\ref{fig:hardware} we show the results 
of our numerical simulation and hardware experiments \cite{esperimenti}.
Fig.~\ref{fig:hardware}(a) shows
the exact energy levels for different $\lambda$ and the energy levels 
obtained via VAGT with $T=10$ discretization steps and $S=100$ measurement shots. 
We see that, in spite of the finite discretization steps, finite measurement shots, 
and imperfect gate implementation, results on the IonQ hardware are very accurate, 
while simulations on Aspen-9 did not converge. We run different experiments on 
Aspen-9, always getting similar outcomes, though numerical simulations with 
Rigetti's decoherence and dephasing times show results comparable with IonQ. 
We believe that the high errors on Rigetti's hardware are possibly due to the qubit 
connectivity, that requires extra compilation steps in order to implement 
the non-local gates required by the VAGT circuits. On the other hand, 
all qubits in IonQ's hardware are fully connected, so better results are expected. 
Indeed, we see in Fig.~\ref{fig:hardware} that the 
accuracy obtained with IonQ hardware is very high. 

\begin{figure}[t]
	\centering
	\includegraphics[width=0.95\linewidth]{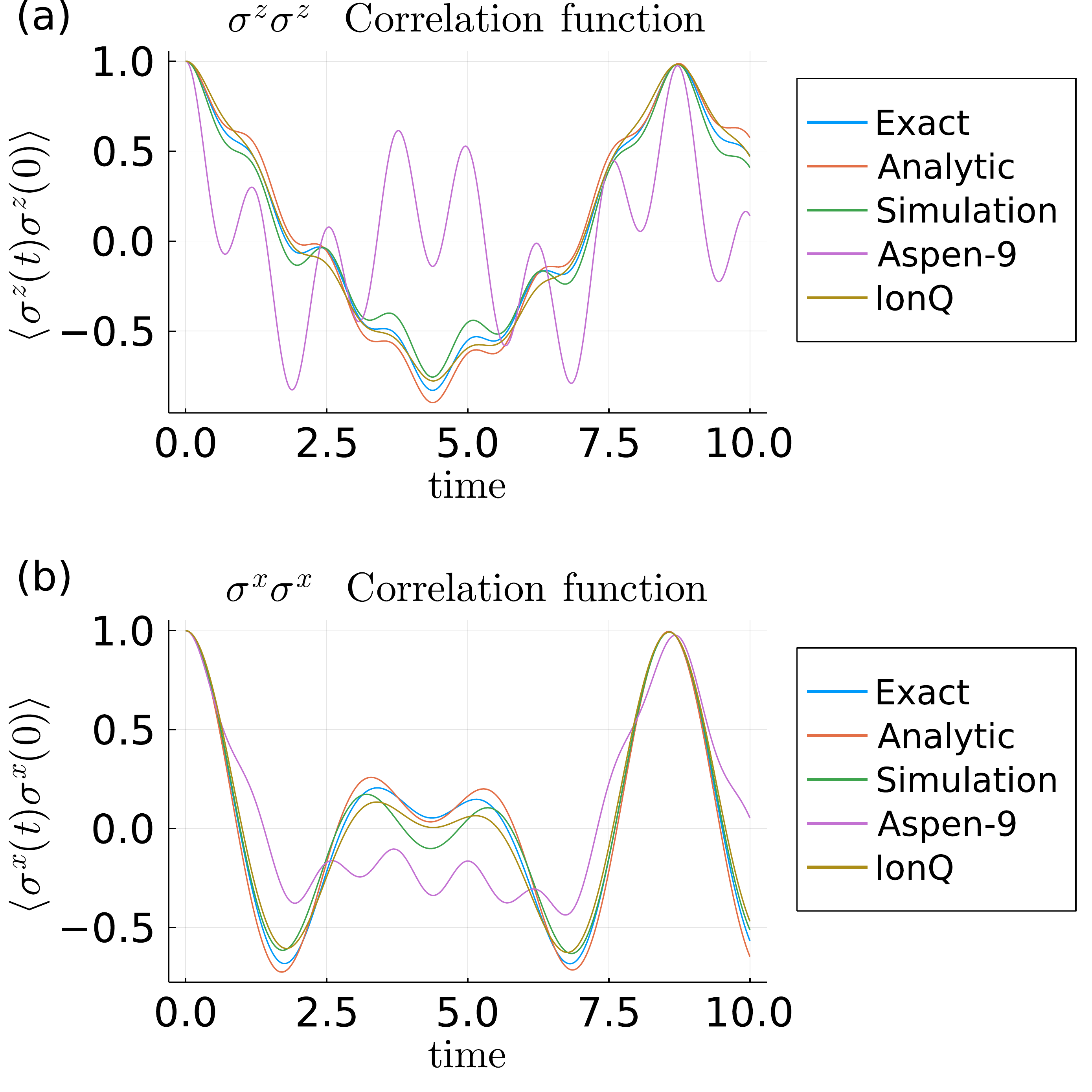}
	\caption{Time-evolved correlation functions, as defined in Eq.~\eqref{eq:correlation}, 
		obtained with different methods, using the Hamiltonian Eqs.\eqref{eq:H0_nondiag_mt}-\eqref{eq:V_ops_mt}.
		The ground states and time evolution operators are either computed exactly (Exact,Analytic), 
		through numerical simulations with shot noise (Simulation), or in the quantum hardware
		(Aspen-9, IonQ). More precisely, in the Exact line we employ
		analytically computed ground states of $H$, obtaining
		$\ket{g_{\text{exact}}}$ and plot $C^{\alpha}_{\text{exact}} = \Re
		\bra{g_{\text{exact}}}\sigma_1^\alpha e^{-it H}
		\sigma_1^{\alpha}\ket{g_{\text{exact}}} $. In all the other cases we use
		instead the correlation function in the form of eq. \eqref{eq:correlation},
		where the AGP is obtained using $T$ steps and $L$ layers, 
		using the same model and parameters of Fig.~\ref{fig:hardware}. 
		The Analytic red line is
		referred to a completely classical simulation of our method, where quantum
		measurements are replaced with analytically computed expectation values,
		the Simulation green line is referred to a classical simulation of our
		algorithm on an ideal quantum processor, where we implemented also a
		quantum measurement process simulation, using $S=100$ shots. Note that an
		ideal quantum processor is, by definition, unaffected by noise. 
		The Aspen-9 and IonQ purple and brown lines show quantities computed from experimental data, 
		obtained using real quantum hardware. 
	}%
	\label{fig:corrfunc}
\end{figure}

In order to study the accuracy of the computed diagonal forms, rather than 
focusing on operator norms or other mathematical distances, we focus on 
the study of physical quantities, like time-evolved correlation functions. 
In Fig.~\ref{fig:corrfunc} we plot the real part of the time-evolved correlation function on the ground state:
\begin{equation}
	C^{\alpha}(t) = \Re \bra{g_0}U^\dagger_\lambda \sigma^{\alpha}_1 e^{-it\tilde{H}_\lambda} \sigma^{\alpha}_1 U_\lambda\ket{g_0}
\label{eq:correlation}
\end{equation}
where $\alpha ={x, z}$, $\ket{g_0}$ is the ground state of $H_0$, so that
$U_\lambda\ket{g_0}$ is our approximation of the ground state of the
Hamiltonian; $\tilde{H}_\lambda$ is the Hamiltonian in its diagonal form,
reconstructed from either experimental or simulation data. 
In Fig.~\ref{fig:corrfunc} we see that all simulations provide an accurate description of the dynamics,
with the exception of the simulations on the Rigetti hardware.

\section*{Discussion}
We have defined the VAGT hybrid quantum algorithm for block- and full-diagonalization 
of many-body Hamiltonians. It can be used to extract low-energy effective theories in 
complex many-particle systems or to approximate long-time evolutions, e.g.~using 
fast forwarding. 

The VAGT is based on the adiabatic gauge potential (AGP), a non-perturbative method 
that generalizes the adiabatic theorem to multiple energy levels. 
The AGP has been successfully used in both analytical calculations with toy models and numerical 
simulations with classical computers, which are nevertheless limited to few-body operators, 
 because of the exponentially large Hilbert space. 
The VAGT algorithm on the other hand is specifically made for hybrid quantum-classical simulations, 
where the complex calculations in exponentially large spaces are efficiently 
performed by the quantum hardware. 
It uses a variational quantum circuit approximation of the unitary transformation generated by the AGP, 
whose optimal parameters are iteratively obtained by merging outcomes from purpose-built
quantum measurements with simple classical post-processing routines. 
When a Hamiltonian can be transformed into a (block) diagonal 
form using a shallow parametric circuit, then the VAGT algorithm can find the optimal 
parameters efficiently, using a number of classical and quantum operations that
scale polynomially in the number of qubits. 

We remind that the algorithm relays on the use of suitable ansatz for the choice of the shallow parametric circuit: How to find such an ansatz, or even understand if it can exist for a given target Hamiltonian, are obviously rather relevant questions, which are out of the scope of the present paper.

To show the performance of the VAGT algorithm, we have considered both random and 
physically motivated Hamiltonians,  where the block structure may come from 
separated energy bands or may be defined by the symmetries of the problem. 
We have performed both exact numerical simulations and simulations with realistic 
error sources (e.g.~measurement shots), 
always obtaining convergence after a few iterations. Moreover, we have also run our algorithm on Rigetti and 
IonQ quantum computers, finding very accurate results on the latter, possibly thanks to its all-to-all qubit connectivity.

\subsection*{Acknowledgements} 
This material is based upon work supported by the U.S. Department of Energy,
Office of Science, National Quantum Information Science Research Centers,
Superconducting Quantum Materials and Systems Center (SQMS) under the contract
No. DE-AC02-07CH11359. L.B.~acknowledges support by the program ``Rita Levi Montalcini''
for young researchers. 

\clearpage


\appendix

\section{Adiabatic Gauge Potential} \label{sec:AGP}
Consider a system evolving under the Hamiltonian $H_{\mu(t)}$ defined in Eq.\eqref{eq:H_lambdadef}, which is time-dependent through the parameter $\mu(t)$ and is written as a matrix in the {\it computational basis} where 
all $\sigma_j^z$ are diagonal.
We define $U_{\mu(t)}$ the instantaneous unitary operator that diagonalizes the Hamiltonian at time $t$ 
\begin{equation}
\tilde{H}_{\mu(t)}=U^\dagger_{\mu(t)} H_{\mu(t)} U_{\mu(t)} \quad \forall t \, ,
\end{equation}
with respect to an eigenbasis of $H_0$,
and we'll always denote with $\sim $ the operators in this reference frame. 
For the moving observer in the instantaneous eigenbasis of $\hat{H}_{\mu(t)}$ the effective Hamiltonian ruling the dynamics is \cite{SelsE3909}: 
\begin{equation}
\hat{H}^{eff}=\tilde{H}_{\mu(t)} - \dot{\mu}\tilde{\mathcal{A}}_{\mu(t)} \, ,
\end{equation}
where $ \tilde{\mathcal{A}}_{\mu(t)} $ is the Adiabatic Gauge Potential (AGP) in the moving frame: 
\begin{equation}
\tilde{\mathcal{A}}_{\mu(t)}= i U^{\dagger}_{\mu(t)}\partial_\mu U_{\mu(t)}\, ;
\end{equation} 
the AGP in the standard frame is: 
\begin{equation}
\label{eq:A_def}
    \mathcal{A}_{\mu(t)}=U_{\mu(t)} \tilde{\mathcal{A}}_{\mu(t)} U^\dagger_{\mu(t)}= i \partial_\mu U_\mu \,  U^\dagger_\mu \, .
\end{equation}
It is possible to show \cite{SelsE3909, Polkovnikov17} that:
\begin{equation}
\label{eq:motion}
    i (\partial_{\mu}H_{\mu} + F_{\text{ad}}) = [ \mathcal{A}_{\mu}, H_{\mu}] \, ,
\end{equation}
where
\begin{equation}
F_{\text{ad}}= \sum_n \left(\partial_{\mu}\epsilon^n_{\mu}\right) \ket{n_{\mu}}\bra{n_{\mu}}
\end{equation}
is the adiabatic, or generalized force, operator \cite{SelsE3909}, $\ket{n_{\mu(t)}}$ being the istantaneous $n$-th eigenstate of $H_{\mu(t)}$.  \\\\

Let's now suppose we want diagonalize the Hamiltonian $H_{\mu(t)}$ at any time $t$. Instead of calculating directly the unitary operator $U_\mu$, we can search for its instantaneous generator, the AGP. We use a \textit{variational} approach, in the sense that we make an hypothesis of a suitable form of $\mathcal{A}_\mu$, $A_{\mu}(\alpha)$, depending on some variational parameters $\alpha$.  \\
Let's now define: 
\begin{equation}
\label{eq:G}
    G_{\mu}(\alpha)= \partial_\mu H_\mu + i [A_{\mu}(\alpha) ,H_\mu ] \, ; 
\end{equation}
from equation \eqref{eq:motion}, if there is a set $\alpha^*$ of variational parameters 
such that $G_\mu = -F_{\text{ad}}$   we also have
\begin{equation}
\label{eq:A_best}
A_{\mu}(\alpha^*)=\mathcal{A}_\mu + D_\mu \, ,
\end{equation}
where $D_\mu$ is a operator that commute with $H_{\mu}$. In other words, if we find the set of optimal variational parameters $\alpha^*$, we find the AGP apart of diagonal elements in the basis of the instantanous eigenstates of the Hamiltonian. \\
In \cite{Polkovnikov17} it is formally demonstrated that searching for the operator $A_\mu$ that minimize the distance from $G_\mu$ and $F_{\text{ad}}$, that is searching for the best approximation of the AGP, is equivalent to finding the variational parameters that minimize $G_\mu (\alpha)$ operator norm: 
\begin{equation}
\label{eq:cost}
\underset{\alpha}{\text{min}} \, \vert \vert { G_\mu (\alpha)} \vert \vert = \underset{\alpha}{\text{min}} \, \vert \vert { \partial_\mu H_\mu + i [A_{\mu}(\alpha) ,H_\mu ]} \vert \vert  \, .
\end{equation}
We remind that solving this equation leads to the best approximation of the AGP, except for the diagonal part, that is undetermined by construction, as we explicitly stated by equation \eqref{eq:A_best}.\\

\section{Our variational ansatz}\label{sec:variational}
Taking inspiration from variational hybrid quantum-classical computation \cite{yuan2019theory, GentiniC20}, we propose to use a ''quantum circuit"-type ansatz of the operator $U_\mu$: 
\begin{equation}
\label{eq:U}
U_\mu(\alpha)= U_0 \prod_{l=1}^{\overset{L}{\rightarrow}} e^{-i \alpha^l_{\mu} B^l} \, ,
\end{equation}
where $L$ is the number of layer in the circuit ansatz, the $B$ operators are one- or two-local operators (loosely speaking: $e^{i \alpha B}$ is a one or two-qubit gate); and the arrow over the product sign define the order of the product itself: specifically: 
\begin{gather}
   \prod_{l=1}^{\overset{L}{\rightarrow}} \mathcal{U}_l := \mathcal{U}_1 \,  \mathcal{U}_2 ... \mathcal{U}_L \\
   \prod_{l=1}^{\overset{L}{\leftarrow}} \mathcal{U}_l := \mathcal{U}_L\, \mathcal{U}_{L-1} ... \mathcal{U}_1 \, .
\end{gather}
Note that in this paper we supposed to know the eigenvalues and eigenstates of the $H_0$ Hamiltonian defined in the main text in equation \eqref{eq:H_lambdadef}, meaning we can efficiently construct the quantum circuit realizing the rotation $U_0$ such that 
\begin{equation}
    D_0 = U_0^{\dagger} H_0 U_0  
\end{equation}
is diagonal, and consequently it is a constant element in the definition of our variational circuit ansatz above. 

We can compute the generator of $U_\mu$, that will be our variational hypotesis for $A_\mu$, from its definition in Eq. \eqref{eq:A_def}: 
\begin{equation}
\label{eq:A_var}
    A_\mu(\alpha)= U_0 \sum_{k=1}^L \Biggl[ \,  \prod_{l<k}^{\rightarrow} e^{-i \alpha^l_{\mu} B^l} \, \frac{d \alpha^k_\mu}{d\mu}\,  B^k \,  \prod_{l<k}^{\leftarrow} e^{i \alpha^l_{\mu} B^l}  \Biggr] U_0^{\dagger} 
\end{equation}

Equations \eqref{eq:U} and \eqref{eq:A_var} are valid for any value of $\mu$. If we want to (block-) diagonalize the Hamiltonian in equation \eqref{eq:H_lambdadef} we can assume that $\mu$ is a running parameter, $\mu \in 0...\lambda$, and iteratively find the generator $A_\mu$ for all points $\mu \in 0...\lambda$. \\

At this level, each parameter $\alpha^l_\mu$ is a continuous function of the running parameter $\mu$. If we now divide the interval $0 \dots \lambda$ in $T$ intervals $\delta\mu$, we create a discrete set of $T$ values for $\mu$: 
\begin{gather}
\mu \in 0 \dots \lambda \rightarrow \{ \mu_t\} _{t=1}^T  \\ 
\mu_t=  t \delta \mu \, , \quad t \in \{ 1 \dots T \} \in \mathbb{N} . \nonumber
\end{gather} 
As a result, we now have a discrete set of variational parameters, $\{\alpha\}_{l, t}^{L,T}$; $U_{\mu_t}$ at the step $t$ is expressed as a parametric evolution, depending on variational parameters $\alpha_t$, and the expression of the generator at the step $t$ is now:
\begin{equation}
\label{eq:A_opt}
A_{\mu_t}(\alpha_t, \alpha_{t+1}) = \sum_{k=1}^{L} \frac{\alpha^k_{t+1}-\alpha^k_{t}}{\delta \mu}O^k_{t} \,,
\end{equation} 
 where we use the finite difference form for the derivative of $\alpha$ \textit{wrt} $\mu$ and we have defined: 
\begin{gather}
\label{eq:Uk} 
U^k_{t} := U_0 \prod_{l<k}^{\rightarrow} e^{-i \alpha^l_{t} B^l}\, , \\
\label{eq:Olt}
O^k_{t} :=U^k_t B^k (U^k_t)^{\dagger}\, .
\end{gather}
Since the Hamiltonian $H_0$ in equation \eqref{eq:H_lambdadef} is diagonalized by $U_0$, we can take advantage of the fact that $U(\mu_t=0)\vert_{\{\alpha\}=0}= U_0$: this means that we know that the optimized parameters $\{\alpha\}_{l, t=0}^{L, N}$ at the step $t=0$ are all zero.\\
Morover, $A_{\mu_t}(\alpha_t, \alpha_{t+1})$ is a function of only the subset of $\{\alpha\}_{l, t}^{L}$ and $\{\alpha\}_{l, t+1}^{L}$ at the step $t$ and $t+1$ respectively.
This two facts leads to an iterative method to solve eq.  \eqref{eq:cost}:
\begin{itemize}
\item first compute eq. \eqref{eq:cost} at the step $t=0$ with $\{\alpha\}_{l, t=0}^{L}=0$, that we know are optimized already ; \item this gives a function of only $\{\alpha\}_{l,t=1}^{L}$, the subset of parameters at $t=1$, that can be easily optimized (see appendix \ref{sec:analytic} for more details);

\item once the optimal parameters for $t=1$ have been obtained, one has to repeat the previous two steps in order to obtain the optimal parameters for $t=2$ and so on, until one reaches the last step $t=T$, that corresponds to $\mu=\lambda$ and the correct generator $A(\lambda)$ for the unitary operator $U(\lambda)$ is finally obtained. 
\end{itemize}

\subsection{Analytic minimization}\label{sec:analytic}
As we said previously, at the time step $t$ we want optimize the cost: 
\begin{equation}
    C_{\mu_t}(\alpha_{t+1}, \alpha_t)=  \vert \vert { \partial_{\mu_t} H_{\mu_t} + i [A_{\mu_t}(\alpha_{t+1}, \alpha_t) ,H_{\mu_t}]} \vert \vert^2 
\end{equation}
\textit{w.r.t.} $\alpha_{t+1}$, since $\alpha_t$'s are optimized already. \\ 
Using the form \eqref{eq:A_opt} and the equation \eqref{eq:H_lambdadef} for the Hamiltonian, we find: 
\begin{align}
\label{eq:an_cost}
    C_{\mu_t}(&\alpha_{t+1}, \alpha_t) = \vert \vert { V + \sum_{l=1}^{L} \frac{\alpha^l_{t+1}-\alpha^l_{t}}{\delta \mu} i [O^l_{t} ,H_{\mu_t} ]} \vert \vert^2  \\
    &= \text{Tr}(V V) +\frac{2}{\delta \mu} \sum_{l=1}^{L} (\alpha^l_{t+1}-\alpha^l_{t}) \text{Tr}(V Q^l_t) + \nonumber \\
    &+ \frac{1}{\delta \mu^2} \sum_{l, \ell}^L (\alpha^l_{t+1}-\alpha^l_{t})(\alpha^{\ell}_{t+1}-\alpha^{\ell}_{t})\text{Tr}(Q^l_t Q^{\ell}_t) \, ,\nonumber
\end{align}
where we use the fact that $V$ is Hermitian and we defined: 
\begin{equation}
\label{eq:Q}
 Q^l_t = (Q^l_t)^\dagger =  i [O^l_{t} ,H_{\mu_t} ] \, .  
\end{equation}

In order to minimize the cost \eqref{eq:an_cost}we can now calculate the gradient and set it to zero: 
\begin{align}
    \frac{\partial C_{\mu_t}(\alpha_{t+1}, \alpha_t)}{ \partial \alpha^{\ell}_{t+1}} & =0 \quad \forall \ell \in 1...L \\
    \label{eq:system}
    \sum_{l=1}^{L} \frac{(\alpha^l_{t+1}-\alpha^l_{t})}{\delta \mu } \text{Tr}(Q^{\ell}_t Q^l_t) & = - \text{Tr}(V Q^{\ell}_l) \quad \forall \ell \in 1...L \, .
\end{align}
The equation \eqref{eq:system} is a linear system that we can solve numerically:
once the solution for
\begin{equation}
\label{eq:beta}
    \beta^l_t \equiv \frac{(\alpha^l_{t+1}-\alpha^l_{t})}{\delta \mu }
\end{equation} 
is found,
we can write the optimized parameters $\alpha^{\ell}_{t+1} \, \, \forall \ell$ as 
\begin{equation}
\label{eq:GD}
\alpha^{\ell}_{t+1} = \alpha^{\ell}_{t} + \delta \mu \beta^{\ell}_t \, .
\end{equation}

\subsection{Quantum circuits}\label{sec:circuits}
For every step $t \in 1...T$ the method involves the solution of the linear system \eqref{eq:system} so, at each step, we need to compute:
\begin{equation}
\label{eq:b}
   b^l= - \text{Tr}(V Q_l)\,,
\end{equation}
\begin{equation}
\label{eq:X}
   X^{\ell, l} = \text{Tr}(Q_{\tilde l} Q_l) 
\end{equation} 
$ \forall \quad \ell, l, \in 1...L$, where we dropped the $t$ step index to simplify the notation. \\
In order to evaluate them through a quantum computer, let's suppose: 
\begin{equation}
    \label{eq:decompose}
    V=\sum_j v_j \sigma_j \quad \quad H_{\mu}=\sum_j h_j \sigma_j \, ,
\end{equation}
where $\sigma_k$ are strings of Pauli operators $\forall \, \,  k$ that forms a complete basis of $SU(N)$ (eventually, some coefficients $v_j$ and $h_j$ may be zero, depending on the specific model). Our quantities become: 
\begin{equation}
     b^l= -\sum_{j, k} v_j h_k \text{Tr}(\sigma_j i[U^l B^l U^{l \dagger}, \sigma_k]) 
\end{equation}
\begin{equation}
   X^{\ell, l} = \sum_{j, k} h_j h_k \text{Tr}(i[U^\ell B^\ell U^{\ell \dagger}, \sigma_j] \,  i[U^{l} B^{l} U^{l \dagger}, \sigma_k]) 
\end{equation} 
where we use also equations \eqref{eq:Olt} and \eqref{eq:Q}, again dropping $t$-labels. \\

We are now left with the task of evaluating on a quantum computer the following two types of terms:
\begin{enumerate}
    \item[1.] $\text{Tr}(\sigma_j \, i[U^l B^l U^{l \dagger}, \sigma_k])$
    \item[2.] $ \text{Tr}(i[U^\ell B^\ell U^{\ell\dagger}, \sigma_j] \,  i[U^{l} B^{l} U^{l \dagger}, \sigma_k])$\,.
\end{enumerate} 
For this aim, we consider the identity below:
\begin{equation}
    \label{eq:iden}
    \text{Tr}(AB)=2^N \bra{\phi} A^T \otimes B \ket{\phi} \, ,
\end{equation}
where $A$ and $B$ are hermitian operators acting on a $2^N$-dimensional Hilbert space $\mathcal{H}$, $T$ indicates the transpose and $\ket{\phi}$ is the maximally entangled state defined as
\begin{equation}
    \ket{\phi}=\frac{1}{\sqrt{2^N}}\sum_{i=0}^{2^N -1} \ket{ii} \quad \in \mathcal{H}\otimes \mathcal{H} \, ,
    \label{eq:phiN}
\end{equation}
and $\{\ket{i} \}$ is a orthonormal basis for $\mathcal{H}$. 
Using \eqref{eq:iden} we can now express our target terms as: 
\begin{enumerate}
    \item[1.] $ 2^N \bra{\phi} \sigma_j^T \otimes i[U^l B^l U^{l \dagger}, \sigma_k] \ket{\phi}$
    \item[2.] $2^N \bra{\phi} (i[U^\ell B^\ell U^{\ell \dagger}, \sigma_j])^T \otimes  i[U^{l} B^{l} U^{l \dagger}, \sigma_k] \ket{\phi}$\,.
\end{enumerate} 
Note that $\sigma_j$ are strings of Pauli operators, so $\sigma_j^T=(-1)^{n_j^y}\sigma_j$, where $n_j^y$ is the number of $\sigma^y$ in $\sigma_j$. Furthermore 
\begin{equation}
    \label{eq:transpose2}
    {(i[U^\ell B^\ell U^{\ell \dagger}, \sigma_j])}^T = -i[{(U^{\ell \dagger})}^T {(B^\ell)}^T {(U^\ell)}^T, \sigma_j^T] \, .
\end{equation}
We can choose a variational ansatz where $(B^\ell)^T \equiv B^\ell \, \forall l$ ( in other words, we can use an ansatz in which $\sigma^y$ appear an even number of times in each operator $B^\ell$), so: 
\begin{equation}
    \label{eq:transpose3}
    (i[U^\ell B^\ell U^{\ell \dagger}, \sigma_j])^T= -i(-1^{n_j^y})[V^{\ell \dagger} B^\ell V^\ell, \sigma_j] \, ,
\end{equation}
where $V$ is our variational ansatz $U$, defined in equation \eqref{eq:Uk}, taken in the reverse order, i.e.: 
\begin{gather}
    U^k := \prod_{l<k}^{\rightarrow} e^{-i \alpha^l B^l}\,, \nonumber\\ 
    V^k=(U_k)^T =  \prod_{l<k}^{\leftarrow} e^{-i \alpha^l (B^l)^T} =  \prod_{l<k}^{\leftarrow} e^{-i \alpha^l B^l} \, ;
\end{gather}
 (of course one can choose an ansatz in which $B^l$ too is not symmetric, in this case a factor $(-1)^{n_l^y}$, where $n_l^y$ is the number of $\sigma^y$ in $B_l$, must be inserted in the exponential in the equation above, as well as in equation \eqref{eq:transpose3}). \\ 

Our target terms can now be written as: 
\begin{enumerate}
    \item[1.] $ 2^N (-1)^{n^y_j}\bra{\phi} \sigma_j \otimes i[U^l B^l U^{l\dagger}, \sigma_k] \ket{\phi}$
    \item[2.] $2^N (-1)^{n_j^y+1} \bra{\phi} i[V^{\ell \dagger} B^\ell V^\ell, \sigma_j] \otimes  i[U^{l} B^{l} U^{l \dagger}, \sigma_k] \ket{\phi}$
\end{enumerate} 

For the first type of terms, we can consider the quantum circuit of Fig.~\ref{fig:circuits}(a),
where $R_x$ is:
\begin{equation}
    \label{eq:Ry}
    R_x=\frac{1}{\sqrt{2}}(\mathbb{I} -i \sigma^x)\, ,
\end{equation} 
and the construction of the state $\ket{\phi}$ from the standard initial state $\ket{00...0}$ is made by applying the Hadamard gate on the first $N$ qubit, followed by $N$ C-NOT gates, controlled by the first $N$ qubit with the second half of the register as a target. In particular, for $N=2$ the quantum circuit is the one of Fig.~\ref{fig:circuits}(b),
that constructs the state $\ket{\phi}$ and can be used as an input to the circuit above.
Once the ancilla qubit is measured in the $\sigma_z$ basis, the probability $p_{A}(0)$ of getting the outcome $0$ is linearly related to the target: 
\begin{equation}
    \label{P_true}
    \bra{\phi} \sigma_j \otimes i[U^l B^l U^{l\dagger}, \sigma_k] \ket{\phi} = 2 - 4 p_A(0) \, .
\end{equation}

Similarly, the circuit of Fig.~\ref{fig:circuits}(c) can be employed to evaluate the second type of terms, as their value are encoded in the probability of getting the outcome $0$ from the first ancilla qubit measurement through the relationship
\begin{equation}
    \bra{\phi} i[V^{\ell \dagger} B^\ell V^\ell, \sigma_j] \otimes  i[U^{l} B^{l} U^{l \dagger}, \sigma_k] \ket{\phi} = -4 + 8p_A(0)\,.
\end{equation}

\section{Saving money: an approach for \texorpdfstring{$N=2$}{TEXT}}
\label{sec:savingmoney}
Even if the algorithm proposed in the previous section is feasible on NISQ devices within reasonable limits, the real costs of running experiments on real devices can be high if a cost is charged for circuit reconfiguration, that is implicit at any step in our variational approach. Consequently, we develop an alternative method to reduce the number of quantum circuits employed by the algorithm itself: Although this method can lead to lower costs (and we actually used it in our experiments), we  remark that it is not efficient from the  point of view of scalability,so that it is practically useful for very small values of $N$ only. 
Dropping again the step index $t$, let's consider the equation \eqref{eq:an_cost}:
\begin{equation}
    C(\alpha_{t+1}, \alpha_t) = \vert \vert { V + \sum_{l=1}^{L} \beta_l  Q_l} \vert \vert ^2 
\end{equation}
where we used definitions \eqref{eq:Q} and \eqref{eq:beta} for $ Q_l$ and $\beta_l$.\\
We can expand the operators $V$ and $Q_l \, \forall l $ on the basis of Pauli strings, obtaining expressions with at most $4^N$ operators (this expansion is exponentially inefficient, but for $N=2$ it leads to a 16-terms expansion, that we can afford easily.): 
\begin{align}
    V &= \sum_{i=0}^{4^N} v_i \sigma_i \\
    Q_k &= \sum_{j=0}^{4^N} q_{k j} \sigma_j 
\end{align}
where $\sigma_i$ is a Pauli string of two operators and, by definition:
\begin{align}
\label{eq:vi}
    y_i &= \frac{1}{2^N}\text{Tr}(V \sigma_i) \\
    \label{eq:qkj}
    q_{kj} &= \frac{1}{2^N} \text{Tr}(Q_k \sigma_j)\,.
\end{align}
After some calculations, recalling that 
\begin{equation}
    \text{Tr}( \sigma_i \sigma_j) = 2^N \delta_{ij} 
\end{equation}
we obtain
\begin{equation}
    \label{eq:costlinreg}
    C(\alpha_{t+1}, \alpha_t)= 2^N ( \mathcal{V} + \mathscr{Q} \beta)^T ( \mathcal{V} + \mathscr{Q} \beta) \, , 
\end{equation}
where $\mathcal{V}$ is the vector of $v_i$'s, $\mathscr{Q}$ is the matrix composed by $q_{kj}$ and $\beta$ is the vector of $\beta_i$'s.
We recall that the cost function, as it's expressed in equation \eqref{eq:costlinreg} is the well-known least-squares loss function of a multiple linear regression classical problem, that can be solved efficiently.\\ 
Therefore, our goal is now to estimate efficiently the elements of $\mathcal{V}$ and $\mathscr{Q}$ defined in eq. \eqref{eq:vi} and \eqref{eq:qkj}, respectively. \\
For what concern $y_i$, they are already known in our setting, since they are coefficients in Pauli decomposition of the operator $V$, the hard-to-diagonalize part of the Hamiltonian. \\
Let's focus on $q_{kj}$. Decomposing 
\begin{equation}
\label{eq:H_dec}
    H_{\mu}=\sum_l h_l \sigma_l 
\end{equation}
and taking into account equations \eqref{eq:Q} and \eqref{eq:iden} the quantity we want to estimate is
\begin{align}
\label{eq:target_new}
q_{kj} &= \frac{1}{2^N} \text{Tr}(Q_k\sigma_j) = \\&=\frac{2^N}{2^N} (-1)^{n_j^y} \sum_l h_l \bra{\phi} \sigma_j \otimes i[U^k B^k U^{k \dagger}, \sigma_l] \ket{\phi} \, ,\nonumber
\end{align}
where $n_j^y$ is the number of $\sigma^y$ Pauli operators in the string $\sigma_j$. We already know that we can estimate this quantity via the first type of quantum circuit presented in the section \ref{sec:circuits}.  \\

\begin{figure}[th]
    \centering
    \includegraphics[width=0.45\textwidth]{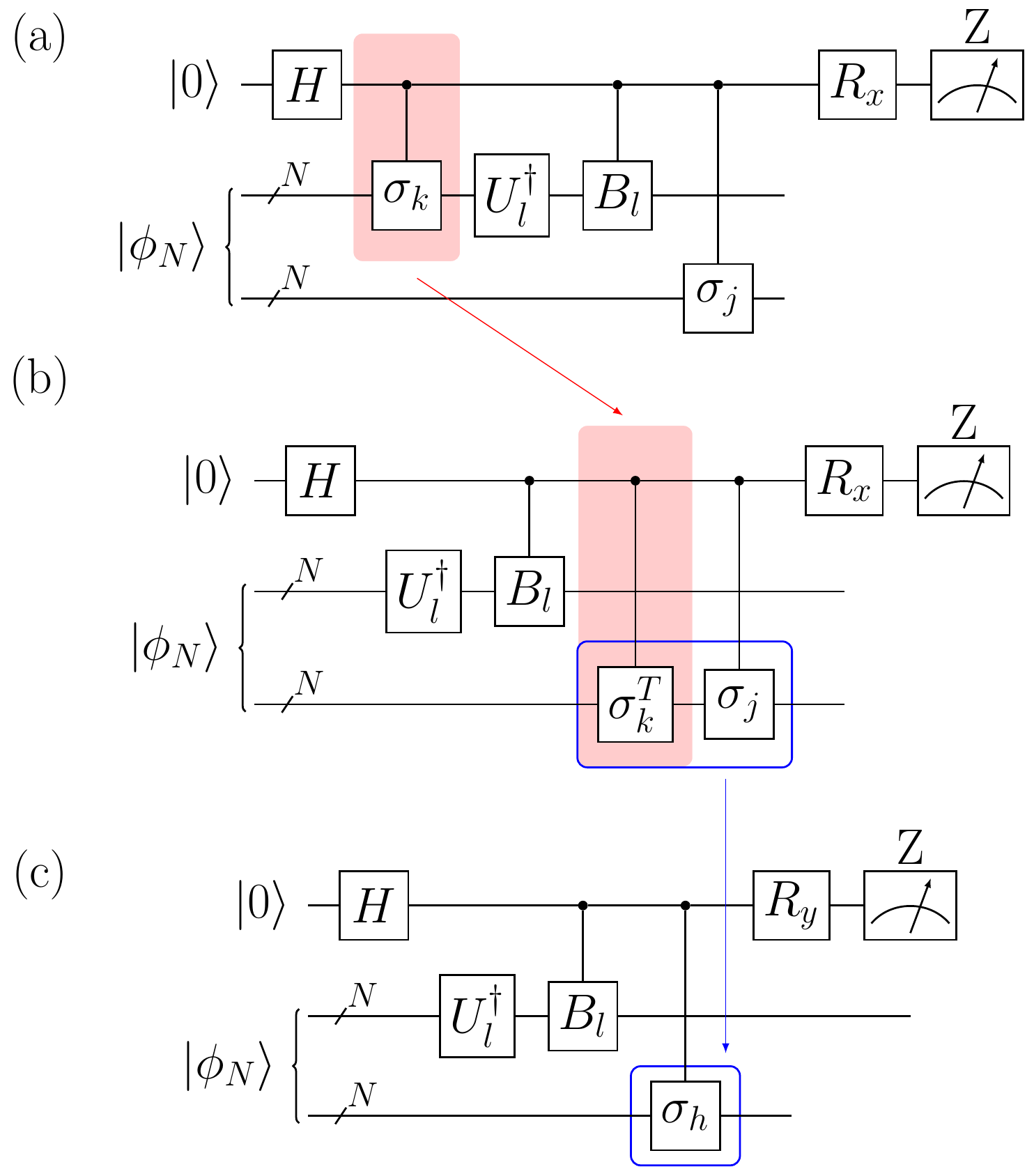}
    \caption{Three quantum circuit carrying the same information: in (a) the original circuit employed in the general method presented in appendix \ref{sec:circuits}; In (b) a completely equivalent circuit, with the same output ; In (c) a further simplification of the circuit, where we used the information about $SU(4)$ algebra's structure coefficients}
    \label{fig:opt_circ}
\end{figure}

Even for $N=2$, the method presented above seems to perform worse than the one presented in section \ref{sec:circuits}. In fact, the number of circuit we have to execute, we have
\begin{equation}
    \mathcal{O}(N^{\gamma}\,  L \,  4^N \,  T) \, , 
\end{equation}  
where $N^{\gamma}$ is the number of terms in the decomposition \eqref{eq:H_dec}, $L$ is the length of the variational ansatz, $T$ is the number of steps in the discretization of the parameter $\mu$ and the highly inefficient factor $4^N$ comes from the decomposition of $Q_k$. \\
On the other hand, by this method we only have quantum circuit of the type shown in Fig \ref{fig:circuits}(a), that can be reduced. In fact, consider the figure \ref{fig:opt_circ}: (a) panel shows the quantum circuit we have to execute in order to calculate $q_{kj}$, while (b) panel shows a completely equivalent quantum circuit. \\
The probabilities of getting the outcome $0$ on the ancilla qubit are respectively: 
\begin{equation}
    \begin{cases}
    P^{ljk}_\text{(a)}(0)=  \frac{1}{2} -\frac{1}{4}\bra{\phi} \sigma_l \otimes i[U^k B^k U^{k \dagger}, \sigma_j] \ket{\phi} \\
    \label{eq:P2}
    P^{ljk}_\text{(b)}(0)=  \frac{1}{2} -\frac{1}{4}\bra{\phi} i[\sigma_l^T, \sigma_j] \otimes U^k B^k U^{k \dagger} \ket{\phi} 
    \end{cases}
\end{equation}
and the equivalence between them follows from identity \eqref{eq:iden}, and from now on we simply denote both of them with $P^{ljk}(0)$.

As we can see also from Eq. \eqref{eq:P2} the second circuit gives a probability connected with the commutator
\begin{equation}
    \Gamma_{lj}= [\sigma^T_l, \sigma_j] = \mathcal{F}_{ljh} \sigma_h
\end{equation}
where $\mathcal{F}_{ljh}$ are the structure coefficients of the algebra.
Using the last equation in the \eqref{eq:P2} we obtain:
\begin{equation}
    P^{ljk}(0)= \frac{1}{2} - \frac{1}{4}i\mathcal{F}_{ljh}\bra{\phi}\sigma_h \otimes U^k B^k U^{k \dagger} \ket{\phi} \, .
\end{equation}
From the equation above it is clear that, although in principle one have to run all different $N^{\gamma} \cdot 4^N$ circuits in fig \ref{fig:opt_circ}(b) with different $\Gamma_{lj}$, since the latter is just a commutator of two sigma strings there are not so many different results for $\Gamma_{lj}$, and so there are not so many different quantum circuits one has to really run. Indeed, one can compute the structure coefficients $\mathcal{F}$ classically and run only 16 ($4^N$ for $N=2$) circuit of the type showed in figure \ref{fig:opt_circ}(c), where $\sigma_h$ is one  of the 16 elements of the $SU(4)$ basis. Keeping trace of the original $lj$-th term corresponding to a given $\sigma_h$ one can recover the information about the original probability $P^{ljk}(0)$. 

Note that, in the quantum circuit in figure \ref{fig:opt_circ}(c) the rotation of the ancilla qubit right before the measurement process is different that in \ref{fig:opt_circ}(a) and (b): 
\begin{equation}
    R_y= \frac{1}{\sqrt{2}} \mathbb{I} - \frac{i}{\sqrt{2}} \sigma_y \, ,
\end{equation}
and the output probability of getting 0 from the ancilla qubit is:
\begin{equation}
\label{eq:P_fake}
    P^{kh}_\text{(c)}(0)= \frac{1}{2} + \frac{1}{2}\bra{\phi}\sigma_h \otimes U^k B^k U^{k \dagger} \ket{\phi}\,,
\end{equation}
and we finally recover the originally searched for probability via
\begin{equation}
    P^{ljk}(0)= \frac{1}{2} -\frac{1}{4}i \mathcal{F}_{ljh} (2P^{kh}_{\text{(c)}} - 1)   \, .
\end{equation}

Finally, we remark that in quantum circuit in fig \ref{fig:opt_circ}(c) it is possible to replace the indirect measurement with a direct one: in other words, it is possible to remove the ancilla qubit, replacing the control-$B_k$ and control $\sigma_h$ gates with measurements of the expectation value of $B_k \otimes \sigma_h$ on the principal register. This is possible only because both $B_k$ and $\sigma_h$ for all $k$ and $h$'s are Pauli strings, so they're observables. 

\section{Full resulting Hamiltonians}
\label{s:Heff}
The full effective Hamiltonian \eqref{eq:Heff} is
\begin{align} \label{eq:fullHeff}
H_{\text{eff}} &\sim -1.10\, \sigma^x_1 + 0.06\, \sigma^y_1 + 1.03\, \sigma^z_1 \\\nonumber 
& -1.10\, \sigma^x_2 + 0.06\, \sigma^y_2 + 1.03\, \sigma^z_2 \\\nonumber & -0.19\, \sigma^x_1 \otimes \sigma^x_2  -0.04\,\sigma^x_1 \otimes \sigma^y_2 \\\nonumber
& -0.03\, \sigma^x_1 \otimes \sigma^z_2  -0.04\,\sigma^y_1 \otimes \sigma^x_2 \\\nonumber
& -0.16\, \sigma^y_1 \otimes \sigma^y_2 + 0.10\,\sigma^y_1 \otimes \sigma^z_2 \\\nonumber
& -0.03\, \sigma^z_1 \otimes \sigma^x_2+ 0.10\,\sigma^z_1 \otimes \sigma^y_2 -0.04\, \sigma^z_1 \otimes \sigma^z_2
\end{align}
where only two decimals are significant, consistently with the choice of $T=100$ and $\delta \mu \sim 1/T$. \\

\bibliography{SW}

\end{document}


\begin{tikzpicture}
    \tikzstyle{every node}=[font=\LARGE]
	\node at (0,0) { 
			\begin{quantikz}
    \lstick{$\ket{0}$} & \gate{H} & \ctrl{1}\gategroup[2,steps=1,style={white,
rounded corners,fill=red!20, inner xsep=2pt}, background]{} & \qw & \ctrl{1} & \ctrl{2} & \qw & \gate{R_x} & \meter{Z}\\
    \lstick[wires=2]{$\ket{\phi_N}$} \qwbundle{$N$}
    & \qwbundle{N} & \gate{\sigma_k} &\gate[wires=1]{U^{\dagger}_l} & \gate{B_l} \qw & \qw & \qw \\
    & \qwbundle{N} & \qw & \qw & \qw & \gate{\sigma_j} \qw  &\qw
    \end{quantikz}
	};
	\node at (0,-6) {
			\begin{quantikz}
    \lstick{$\ket{0}$} & \gate{H} & \qw & \ctrl{1} & \ctrl{2}\gategroup[3,steps=1,style={white,
rounded corners,fill=red!20, inner xsep=2pt}, background]{} & \ctrl{2} & \gate{R_x} & \meter{Z}\\
    \lstick[wires=2]{$\ket{\phi_N}$} \qwbundle{$N$}
    & \qwbundle{N} &\gate[wires=1]{U^{\dagger}_l} & \gate{B_l} \qw & \qw & \qw  & \qw\\
    & \qwbundle{N} & \qw & \qw & \gate{\sigma^T_k}\gategroup[1,steps=2,style={blue,
rounded corners, inner xsep=2pt}, background]{} &\gate{\sigma_j} \qw  &\qw
    \end{quantikz}
	};
    \draw [red, -{latex}](-2,-1.5) -- (1,-3.5);
    
    \node at (0,-12) { 
			\begin{quantikz}
    \lstick{$\ket{0}$} & \gate{H} & \qw & \ctrl{1} & \ctrl{2} & \gate{R_y} & \meter{Z}\\
    \lstick[wires=2]{$\ket{\phi_N}$} \qwbundle{$N$}
    & \qwbundle{N} &\gate[wires=1]{U^{\dagger}_l} & \gate{B_l} \qw & \qw & \qw  & \qw\\
    & \qwbundle{N} & \qw & \qw & \gate{\sigma_h}\gategroup[1,steps=1,style={blue,
rounded corners, inner xsep=2pt}, background]{} \qw  &\qw
    \end{quantikz}
	};
	\draw [blue, -{latex}](2.5,-8.8) -- (2.5,-12.8);
	
	\node at (-7.5,2) {(a)};
	\node at (-7.5,-3.) {(b)};
	\node at (-7.5,-10.2) {(c)};
\end{tikzpicture}